\documentclass[twocolumn,epsf,apl,graphics,psfig,superscriptaddress]{revtex4}

\usepackage{graphicx,graphics} 
\usepackage[dvips,bookmarks=false]{hyperref}
\usepackage{graphicx}
\usepackage{dcolumn}
\usepackage{eucal}
\usepackage[dvips]{epsfig}

\begin{document}

\title{A spin-filter device based on armchair graphene nanoribbons}
\author{A. Saffarzadeh} \email{a-saffar@tpnu.ac.ir}
\affiliation{Department of Physics, Payame Noor University,
Nejatollahi Street, 159995-7613 Tehran, Iran}
\affiliation{Computational Physical Sciences Laboratory,
Department of Nano-Science, Institute for Research in Fundamental
Sciences (IPM), P.O. Box 19395-5531 Tehran, Iran}
\author{R. Farghadan} \affiliation{Department of Physics, Tarbiat Modares
University, P.O. Box 14115-175 Tehran, Iran }
\date{\today}

\begin{abstract}
The coherent spin-polarized electron transport through a
zigzag-edge graphene flake (ZGF), sandwiched between two
semi-infinite armchair graphene nanoribbons, is investigated by
means of Landauer-Buttiker formalism. To study the edge magnetism
of the ZGF, we use the half-filled Hubbard model within the
Hartree-Fock approximation. The results show that the junction
acts as a spin filter with high degree of spin polarization in the
absence of magnetic electrodes and external fields. By applying a
gate voltage the spin-filtering efficiency of this device can be
effectively controlled and the spin polarization can reach values
as high as 90\%.
\end{abstract}
\maketitle

Graphene nanoribbons (GNRs) and graphene junctions are good
candidates for electronic and spintronic devices due to high
carrier mobility, long spin-relaxation times and lengths, and
spin-filtering effect \cite{Sanvito,Kim,Son}. In fact, the
conduction electrons in carbon-based materials can move very long
distances without scattering due to their small spin-orbit
coupling and low hyperfine interaction. For instance, in GNRs with
zigzag edges, electronic transport is dominated by edge states
which have been observed in scanning tunneling microscopy
\cite{Kobayashi}. These states are expected to be spin-polarized
and make zigzag-edge GNR (ZGNR) junctions attractive for nanoscale
spintronic applications such as spin filters
\cite{Jia,Guo1,Guo2,Kumazaki,Hancock,Lu,Ozaki,Zeng}. In order to
achieve a spintronic device, it is very important to find
nonmagnetic materials where a spin-polarized current can be
injected and flowed without becoming depolarized. The ground state
of ZGNRs has an antiferromagnetic spin configuration where the
total spin ($S$) is zero. However, when the system has different
number of A- and B-sublattice sites, the total spin of the ground
state is $2S=N_A-N_B$ \cite{Lieb} and by appropriate designing,
one can form a ferromagnetic spin configuration at the zigzag
edges. Most of the previous studies on spin-filtering effects have
been focused on the junctions, which consist of ZGNR electrodes.

In this letter, a GNR junction which operates as a spin-filter in
the absence of an external electric field is presented and the
influence of edge atoms on spin transport through the junction
will be examined. We also investigate the sensitivity of the spin
polarization to the gate voltage to obtain a maximum value for
this polarization. An interesting feature of our system is that,
in spite of the previous studies, the type of electrodes in this
junction is armchair-edge GNR and the central part of the system
is a zigzag-edge graphene nanodisk \cite{Rossier,Guo1}, as shown
in Fig. 1(a). In this junction, the right GNR electrode with a
ribbon index $n$=8 has a width $W_R=$0.86 nm, while the left
electrode with a ribbon index $n$=33 has a width $W_L=$3.93 nm.
Furthermore, the central region (nanodisk) with a trapezoidal
shape consists of $N_C=240$ carbon atoms and produces a
ferromagnetic spin configuration at its edges.

\begin{figure}
\centerline{\includegraphics[width=0.9\linewidth]{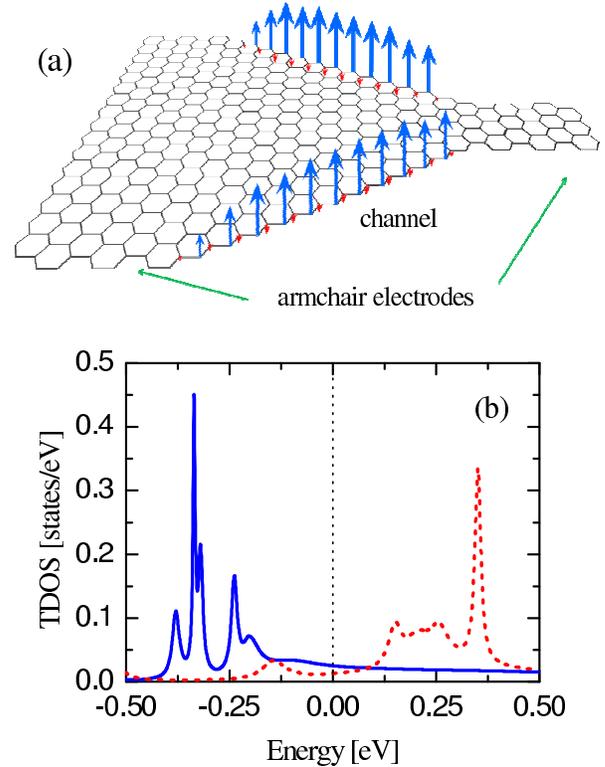}}
\caption{(Color online) (a) The spin-filter device. The device
shows that the localized magnetic moments only form on the edge
atoms of the channel and the electrodes are nonmagnetic. The
upward (blue) arrows correspond to the majority spins, while the
downward (red) arrows correspond to the minority spins. (b) The
majority (solid line) and minority (dashed line) spin densities at
the channel. The vertical dotted line indicates the Fermi energy.}
\end{figure}

We simulate the system depicted in Fig. 1(a) by use of a
single-band tight-binding model and an on-site Hubbard repulsion
treated in the mean-field approximation at half-filling to account
for electron-electron interaction in the junction and calculate
the localized moments on the zigzag-edge atoms. Thus, the mean
field Hamiltonian can be written as \cite{Fujita}
\begin{equation}\label{4}
\hat{H}_{C}=\sum_{i,j,\sigma}(\epsilon_i\delta_{i,j}-t)\hat{d}_{i\sigma}^\dag\hat{d}_{j\sigma}
+U\sum_{i,\sigma}\langle\hat{n}_{i,-\sigma}\rangle[\hat{n}_{i,\sigma}-\frac{1}{2}\langle
\hat{n}_{i,\sigma}\rangle]\ ,
\end{equation}
where $\hat{d}_{i\sigma}^\dag$ and $\hat{d}_{i\sigma}$ are the
electron creation and annihilation operators, respectively, and
$\hat{n}_{i,\sigma}=\hat{d}_{i\sigma}^\dag\hat{d}_{i\sigma}$ is
the number operator for an electron with spin $\sigma$ at site
$i$. Here, $\epsilon_i$ will be set to zero except in the gated
region (channel) where it is equal to gate potential $V_G$.
$t$=2.66 eV is the transfer integral between all the nearest
neighbor sites and $U=1.06\,t$ is the on-site Coulomb interaction.
Starting from an antiferromagnetic configuration as an initial
condition and considering the effect of the semi-infinite GNRs
\cite{Sancho} on the connected atoms of the channel, we solve the
mean-field Hamiltonian self-consistently by iteration method.
Therefore, the Green's function and the spin density on each atom
of the channel should be calculated iteratively until a
convergence of the spin density is reached. The Green's function
of the nanodisk is expressed as
\begin{equation}
\hat{G}_{C}(\omega)=[(\omega+i\eta)\hat{I}-\hat{H}_{C}-\hat\Sigma_{L}-\hat\Sigma_{R}]^{-1}\
,
\end{equation}
where $\eta$ is a positive infinitesimal and $\Sigma_{L,R}$ are
the self-energy matrices due to the connection of left and right
GNR electrodes to the channel. The spin-dependent density of
states and the expectation value for the number operator of
electron on each site of the channel are given as
\begin{equation}
g_{i\sigma}(\omega)=-\frac{1}{\pi}\,\mathrm{Im}\,\langle{i\sigma}\textemdash\hat{G}_{C}(\omega)\textemdash{i\sigma}\rangle\
,\nonumber
\end{equation}
\begin{equation}\label{n}
\langle{\hat{n}}_{i\sigma}\rangle=\int_{-\infty}^{E_F}{g_{i\sigma}(\omega)}d{\omega}\
.\end{equation}

Accordingly, the spin at each site of the channel can be expressed
as $S_i=\frac{\langle \hat{n}_{i\uparrow}\rangle-\langle
\hat{n}_{i\downarrow}\rangle}{2}$. In the mean-field level, there
is not a spin-flip scattering or any other interactions.
Therefore, the spin-polarized transport through such a junction
can be considered within the coherent regime \cite{Datta}. In the
coherent transport, the spin-dependent conductance at low
temperature can be written as
$G_\sigma=\frac{e^2}{h}\mathrm{Tr}[\hat{\Gamma}_{L}
\hat{G}_{C}\hat{\Gamma}_{R}\hat{G}_{C}^{\dagger}]_{\sigma}$, where
$\hat\Gamma_\alpha$, the coupling matrices, can be expressed as
$\hat\Gamma_\alpha(\omega)=-2\,\mathrm{Im}[\hat\Sigma_{\alpha}(\omega)]$.

The total density of states (TDOS) per site for each spin subband,
$\frac{1}{N_C}\sum_{i=1}^{N_C} g_{i,\sigma}(\omega)$, in the
central region and in the presence of the electrodes has been
shown in Fig. 1(b). The spin-dependent electronic states inside
the energy window are completely separated into majority and
minority spin subbands and in the other energy ranges (not shown)
are approximately degenerate. The difference between the two spin
subbands explains the localized magnetization in the channel
edges. In such a case, the total spin of the channel reaches
$S=2.82$, which has a maximum value $S=0.13$ in the middle of the
channel edges [shown in Fig. 1(a)]. We see that the TDOS for
spin-up electrons is more localized at lower energies than the
spin-down ones. The separation between these two spin subbands
which provide two paths for electron conduction through the
junction well explains the spin-filter effect of the system. In
fact, the importance of designing such a device is that the system
at ground state is spin-polarized without applying external
fields.

It has been well known that an armchair edge does not induce local
edge magnetism \cite{Fujita}. Therefore, the electron conduction
through the armchair-edge electrodes is not spin-polarized, but
when the electron arrives at the nanodisk region, different
scatterings occur for the electronic waves with different spin
densities. Therefore, the electron conduction for both spin-up and
spin-down electrons is different at certain energies and the
generation and manipulation of the spin-polarized current would be
possible. We should note that because of the special design for
this junction there is not a significant difference between
ferromagnetic and antiferromagnetic configurations, both for the
electronic structure and the magnetic moment of the atoms on the
edge of the ZGF, hence the transport properties for two spin
configurations are the same.

\begin{figure}
\centerline{\includegraphics[width=0.85\linewidth]{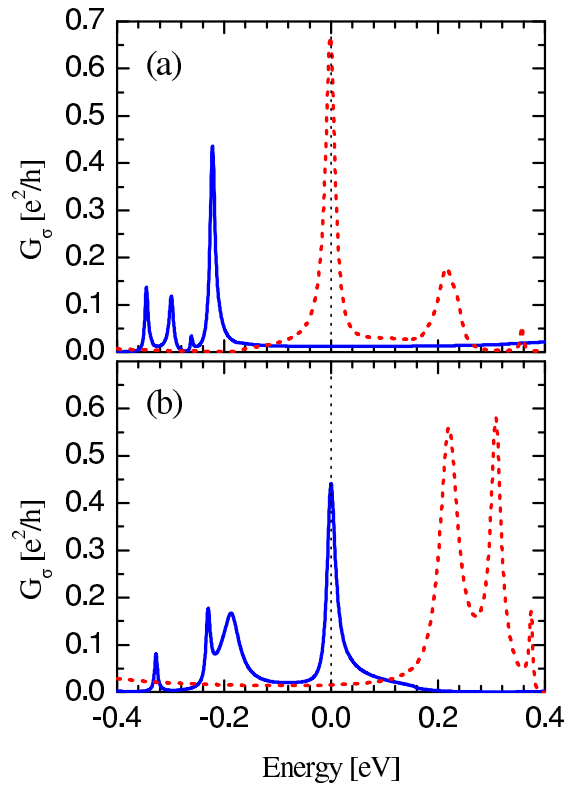}}
\caption{(Color online) (a) Spin-dependent conductance as a
function of energy at (a) $V_G$=\,0.36 V and (b) $V_G$=\,-1.82 V.
The solid (dashed) line is for majority (minority) spin electrons.
The vertical dotted line indicates the Fermi energy.}
\end{figure}

The gate control of spin conduction is an important feature, which
can clearly demonstrate high spin-filter effect in our device and
perhaps in future all-carbon circuits \cite{Zeng}. Applying a gate
voltage changes the potential energy in atomic sites and
accordingly, the electronic states are shifted and slightly
modified (due to the electrodes) by this effect. Hence, the
transmission channels in the system may significantly vary
\cite{Datta,Saffar1}. In a ZGNR, a gate potential may shift the
spin-up and spin-down states differently and therefore, spin
filters and spin switches can be achieved without magnetic
contacts. \cite{Guo1,Saffar2,Lakshmi}. For this purpose, we have
shown in Fig. 2 the spin-dependent conductance as a function of
energy for majority and minority electrons at two different gate
voltages $V_G$=\,0.36 V and $V_G$=\,-1.82 V. We have chosen these
values because the spin-filtering effect in the system can be
easily demonstrated. We can see that the conductance spectrum of
two spin subbands is approximately separated in all electronic
states within the energy window, same as the TDOS shown in Fig.
1(b). At the positive gate voltage ($V_G$=\,0.36 V), the
transmission channel of minority spin electrons is fully open at
the Fermi energy, while that for majority spin electrons is
effectively blocked. Such a feature will be reversed if the
negative gate voltage ($V_G$=\,-1.82 V) is applied.

\begin{figure}
\centerline{\includegraphics[width=0.86\linewidth]{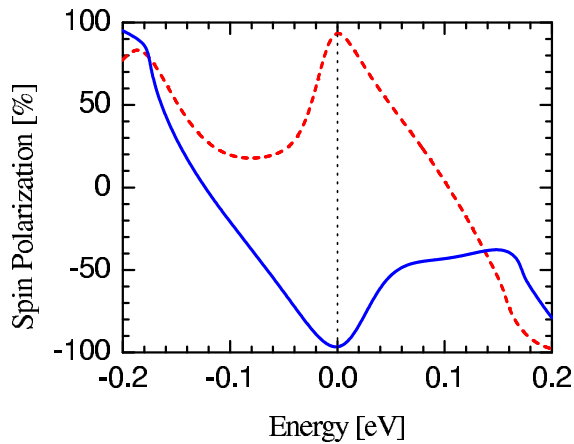}}
\caption{(Color online) (a) Degree of spin polarization as a
function of energy at two different gate voltages. The solid line
is for $V_G$=0.36 V and the dashed line is for $V_G$=-1.82 V.}
\end{figure}

To obtain the spin-filtering efficiency, we have shown in Fig. 3
the degree of spin polarization for electrons traversing the
channel which can be defined as
$P=\frac{G_\uparrow(\omega)-G_\downarrow(\omega)}{G_\uparrow(\omega)+G_\downarrow(\omega)}$.
In this figure, the dip at the Fermi energy clearly represents
that the minority spin electrons at $V_G$=0.36 V carry most of the
current, while the peak is related to the majority spin electrons
at $V_G$=-1.82 V. This can be easily understood from comparing the
behavior of spin-dependent conductance of majority and minority
spin electrons. As seen in Fig. 2, when majority (minority) spin
electrons show high peaks in their conductance spectrum, the
conduction of minority (majority) electrons approximately
vanishes, i.e. high degree of spin polarization. The spin
polarization just like the conductance spectra is sensitive to
changes of energy and in some energy ranges there are high spin
polarizations for each spin subband. We should note that, with
considering the more number of carbon atoms in the channel the
total magnetic moment increases and the spin-filter efficiency is
enhanced. Also, the spin density of each site is calculated at
zero temperature and with considering the effect of temperature in
the number operator Eq.(\ref{n}) \cite{Guo1} the magnetic moment
at the ZGF reduces and this reduction is able to change the
efficiency of this spin filter device. Perhaps the most
interesting aspect of our system is the ability for generation of
spin-polarized currents in the absence of traverse fields and
using electrodes that are nonmagnetic.

In summary, based on the non-equilibrium Green function technique
and the mean-field Hubbard model, we proposed a spin filter device
which preferably passes only one type of spin currents through the
system. In such a device, the current is spin-polarized due to the
finite-size effect and geometry of ZGF. The spin-filtering
efficiency could be controlled by a gate voltage so that the spin
polarization can be higher than 90\% which shows that the system
acts as a spin switch.

The authors would like to thank J. J. Palacios for helpful
discussions.

\end{document}